\documentclass[12pt]{article}
\usepackage{graphicx}      
\def\RR{\vbox {\hbox to 8.9pt {I\hskip-2.1pt R\hfil}}}
\def\NN{{\rm I\hskip-2pt N}}
\def\Gc{{\cal {G}}_c}	 
\def\ds{{\displaystyle}}
\def\CC{{\rm C\hskip-4.8pt \vrule height 6pt width 12000sp\hskip 5pt}}
\def\arg{ x^2/ (4\, t)}

\def\e{{\rm e}}
\def\e{\hbox{e}}

\def\ds{\displaystyle}
\def\RR{\vbox {\hbox to 8.9pt {I\hskip-2.1pt R\hfil}}}
\def\NN{{\rm I\hskip-2pt N}}
\def\CC{{\rm C\hskip-4.8pt \vrule height 6pt width 12000sp\hskip 5pt}}

\def\pni{\par\noindent}
\def\vsh{\smallskip}

\def\vsp{\vsh\pni} 

\DeclareGraphicsExtensions{.eps,.pdf,.jpg,.png}

\begin{document}

\font\title=cmbx12 scaled\magstep2
\font\bfs=cmbx12 scaled\magstep1
\font\little=cmr10
\begin{center}
{\title Propagation Speed of the Maximum of }
\\[0.20truecm]
{\title the Fundamental Solution to}
\\ [0.20truecm]
{\title the Fractional Diffusion-Wave Equation}\footnote{
This paper has been presented by F. Mainardi
  at the International Workshop: FRACTIONAL DIFFERENTIATION AND ITS APPLICATIONS (FDA12) Hohai University, Nanjing, China, 14-17 May 2012 
({\tt http://em.hhu.edu.cn/fda12}).
The peer-revised version of this paper is published in {\it Computers and Mathematics with Applications} {\bf 66}, 774--784 (2013).
[DOI:10.1016/j.camwa.2013.01.005]
The current document is an e-print which differs in e.g. pagination, reference numbering and other typographic details.}
 \\  [0.25truecm]
Yuri LUCHKO$^{(a)}$, Francesco MAINARDI$^{(b)}$ and
Yuriy POVSTENKO$^{(c)}$
\\ [0.10truecm]
$\null^{(a)}${\little Department of Mathematics,}
\\
{\little Beuth Technical University of Applied Sciences, Berlin,
13353 Germany}
\\ {\little E-mail: luchko@beuth-hochschule.de}
\\  [0.10truecm]
$\null^{(b)}$ {\little Department of Physics, University of Bologna, and INFN}
\\
{\little Via Irnerio 46, I-40126 Bologna, Italy}
\\{\little E-mail: francesco.mainardi@unibo.it; francesco.mainardi@bo.infn.it}
\\[0.10truecm]
$\null^{(c)}$ {\little Institute of Mathematics and Computer Science,}
 \\ {\little Jan Dlugosz University in Czestochowa, Czestochowa, 42-200 Poland}
 \\{\little E-mail: j.povstenko@ajd.czest.pl}
\end{center}
\begin{abstract}                
\noindent 
In this paper,  the one-dimensional time-fractional diffusion-wave equation with the fractional derivative of order $\alpha,\ 1 < \alpha < 2$ is revisited. This equation interpolates between the diffusion and the wave equations that behave quite differently regarding their response to  a localized disturbance: whereas the diffusion equation describes a process,  where a
disturbance spreads infinitely fast, the propagation speed of the disturbance is a constant for the wave equation. For the time-fractional diffusion-wave equation, the propagation speed of a disturbance is infinite, but its fundamental solution possesses a maximum that disperses with a finite speed. In this paper, the fundamental solution of the Cauchy problem for the time-fractional diffusion-wave equation, its maximum location, maximum value, and other important characteristics are investigated in detail. To illustrate analytical formulas, results of numerical calculations and plots are presented. Numerical algorithms and programs used to produce plots are discussed.
\end{abstract}

\vsp
{\it Key Words and Phrases}:
Time-fractional diffusion-wave equation, Cauchy problem,  fundamental solution, Mittag-Leffler function
Wright function, Mainardi function.
\vsp  
{\it MSC 2010}: 26A33, 33C47, 33E12, 34A08, 35E05, 
35R11, 44A20, 65D20.


\section{Introduction}
Evolution equations related to phenomena intermediate between diffusion and wave propagation have attracted
the attention of a number of researchers since the 1980's. 
This kind of phenomenon is known to occur in viscoelastic media that combine the characteristics 
of solid-like materials that exhibit wave propagation and fluid-like materials that support diffusion processes. 
In particular, analysis and results presented in \cite{Pipkin_BOOK86}  
and \cite{Kreis-Pipkin_86} should be mentioned. 
Being unaware of an interpretation of evolution equations by means of fractional calculus, these authors
still could provide  an interesting example of the relevance of the intermediate phenomena for models 
in continuum mechanics.
\vsp
Nowadays it is well recognized that evolution
equations can be interpreted as  differential equations of fractional order in
 time when some  hereditary mechanisms of power-law
type are present in  diffusion or wave phenomena. 
This has been  shown for example  
in \cite{Chen-Holm_JASA03,Chen-Holm_JASA04, Mainardi-Tomirotti_GEO97} 
and more  recently in \cite{Mainardi_BOOK10} and \cite{Nasholm-Holm_FCAA13}, where 
 propagation of pulses in linear lossy  
 media governed by constitutive equations of fractional order has been revisited. 
\vsp
For  analysis of the evolution equations of the type mentioned above, 
methods and tools of fractional calculus, integral transforms, and higher transcendental 
functions have been employed in  
the pioneering papers  
\cite{Wyss_JMP86}, \cite{Schneider-Wyss_JMP89}, 
\cite{Fuj90},  
\cite{Gorenflo-Rutman_TMSF94},
\cite{Kochubei_89,Kochubei_90},
and in the book \cite{Prusse_BOOK93}.   
We also mention the papers 
\cite{Mai94,Mai96a,Mai96b} and \cite{Mainardi-Tomirotti_TMSF94},
where fundamental solutions of the evolution equations related to phenomena intermediate between diffusion and wave propagation have been expressed in terms of some auxiliary functions of the Wright type that 
sometimes are referred to as  Mainardi functions, see i.e. \cite{Pod99},  \cite{Gor99,Gor00}. These functions as well as some techniques and methods of fractional calculus, integral transforms, and higher transcendental functions will be used in our analysis.
\vsp
It is well known that diffusion and wave equations behave quite differently regarding their
response to a localized disturbance: whereas the diffusion equation describes a process, where a
disturbance spreads infinitely fast, the propagation speed of the disturbance is constant for the
wave equation. 
In a certain sense, the time-fractional diffusion-wave equation interpolates between these two different responses.  
On the one hand, the support of the solution to this equation is not compact on the real line for each $t>0$ 
for a non-negative disturbance that is not identically equal to zero, i.e. its response to a localized
disturbance spreads infinitely fast (see \cite{Fuj90}). 
 On the other hand, the fundamental solution of the time-fractional
diffusion-wave equation possesses a maximum that disperses with a finite speed similar to the behavior
of the fundamental solution of the wave equation. 
The problem to describe the location of the maximum of the fundamental solution of the Cauchy problem
for the one-dimensional time-fractional diffusion-wave equation of order $\alpha,\ 1<\alpha < 2$ was considered  for the first time in \cite{Fuj90}. Fujita proved that the fundamental solution takes its maximum
 at the point $x_{*}=\pm c_\alpha t^{\alpha/2}$ for each $t>0$, where $c_\alpha >0$ is a constant determined
 by $\alpha$. 
 Recently,  another proof of this formula for the maximum location along with numerical
results for the constant $c_\alpha$ for $1 < \alpha < 2$ were presented in \cite{PovXX}. 
\vsp
In this paper, we provide an extension and consolidation of these results along with some new analytical formulas, 
numerical algorithms, and pictures.
The rest of the paper is organized as follows. 
\vsp
In the 2nd  section, problem formulation and some analytical results are given. 
Here we revisit the results of Fujita and Povstenko and give some new insights into the problem. 
Especially the role of the symmetry group of scaling transformations of the 
time-fractional diffusion-wave equation in the maximum propagation problem is emphasized. 
We derive a new formula for the maximum value of the Green function for the Cauchy problem 
for the time-fractional diffusion-wave equation. 
A new characteristic of the time-fractional diffusion-wave equation - the product of the maximum location 
of its fundamental solution and its maximum value -  is introduced. 
For a fixed value of $\alpha,\ 1\le \alpha \le 2$, 
this product is a constant for all $t>0$ that depends only on $\alpha$.  
The product is equal to zero for the  diffusion equation  and to infinity for the wave equation, 
whereas it is finite, positive, and laying between these extreme values for the time-fractional 
diffusion-wave equation that justifies the fact that the time-fractional 
diffusion-wave equation interpolates between the diffusion and the wave equations. 
The 3rd section is devoted to a presentation of the numerical algorithms used to calculate the fundamental 
solution and its important characteristics including the location of its maximum, 
its propagation speed, and the maximum value. 
Results of numerical calculations and plots are presented and discussed in detail.

\section{Problem formulation and analytical results}

This section is devoted to the problem formulation and some important analytical results. 
In particular, several representations of the fundamental solution of the Cauchy problem 
for the time-fractional diffusion-wave equation  in the form of series and integrals are given. 
These representations are used to derive explicit formulas for the maximum location, maximum value, 
and the propagation speed of the maximum point. 
Besides, we give a new proof of the fact that a response of the time-fractional diffusion-wave equation 
to a localized disturbance spreads infinitely fast like in the case of the diffusion equation. 

\subsection{Problem formulation}

In this paper, we deal with the family of
evolution equations  obtained from the standard
diffusion equation (or the D'Alembert  wave equation)
by replacing  the	first-order (or the second-order) time derivative
 by a fractional  derivative 
of order  $\alpha$ with $1 \le \alpha \le 2$, namely
\begin{equation}
\frac{\partial ^{\alpha}u}{\partial t^{\alpha}}= \frac{\partial ^{2}u}{\partial x^{2}},
\label{eq1}
\end{equation}
where  $x\in \RR $, $t\in \RR^+$ denote the space and time
variables, respectively.
\vsp
In (\ref{eq1}), $u=u(x,t)$ represents the response field variable and the
fractional derivative of order $\alpha,\  n-1 < \alpha < n,\ n\in \NN$ is defined in the Caputo sense:
\begin{equation}
\frac{\partial^{\alpha} u}{\partial t^{\alpha}} =
\frac{1}{\Gamma(n-\alpha)}
 \int_{0}^{t} (t-\tau)^{n-\alpha -1}\frac{\partial^{n}u(\tau)}{\partial \tau^{n}}
 \, {d}\tau,
 \label{eq2}
\end{equation}
where $\Gamma$  denotes the Gamma function. For $\alpha=n,\ n\in \NN$, the Caputo fractional derivative is defined as the standard derivative of order $n$.
\vsp
In order to  guarantee existence and uniqueness of a
solution, we must add to (\ref{eq1}) some initial and boundary conditions.
Denoting by
$f(x)\,,\, x\in \RR\,$ and $\,h(t)\,,\, t\in \RR^+\,$
sufficiently
well-behaved functions,    the Cauchy problem for  the time-fractional diffusion-wave equation with $1\le \alpha \le 2$ is formulated as
follows:
\begin{equation}
\cases{
u(x,0) =f(x) \,,	\ \ -\infty <x < +\infty\,;\cr
u(\mp \infty,t) = 0\,,\ \ \, t>0\,.}
  \label{3.1a}
\end{equation}
If $1 <\alpha \le 2\,, $ we must add
 to (\ref{3.1a}) 
the initial value of the first time derivative of the field variable,
$u_t(x,0)\,,$ since in this case
the  Caputo fractional derivative is expressed in terms
of the second order  time derivative. To ensure continuous dependence  of the solution
with respect to the parameter $\alpha $
we agree  to  assume
$$u_t(x,0) = 0\,, \ \ \hbox{for} \ \ 1<\alpha\le 2\, .$$
\vsp
In view of our subsequent analysis we find it convenient to set
$  \nu  :={\alpha / 2}$, so that 	$1/2 \le \nu \le 1$ for $1\le \alpha \le 2$.
\vsp
For the  Cauchy problem, we introduce
the so-called Green function $\Gc (x,t;\nu )$, 
which represents the respective fundamental solution,
obtained when $f(x) = \delta (x)$, 
$\delta$ being the Dirac $\delta$-function.
As a consequence, the solution of the Cauchy problem
is obtained by a space 
convolution  according to
$$
u(x,t;\nu)
= \int_{-\infty}^{+\infty} \Gc(x-\xi ,t;\nu  ) \, f(\xi)
\, d\xi     \,.
$$
It should be noted that  $\Gc(x ,t ;\nu) =
  \Gc(|x|,t ;\nu)$ since the Green function of the Cauchy problem turns
out to be an even function of $x$. 
This means that we can restrict our investigation of the function $\Gc$ to non-negative values $x\ge 0$. 
\vsp
For the  standard diffusion equation ($\nu   =1/2$) it is well known that
\begin{equation}
\label{diff}
\Gc(x,t;1/2) := \Gc^d (x,t)
 = {t^{-1/2}\over 2\sqrt{\pi }}\,\e^{-\ds \arg}\,.
\end{equation}
\vsp
In the limiting case $\nu  =1$ we recover  the standard wave equation,
for which we get
\begin{equation}
\label{wave}
\Gc(x,t;1) := \Gc^w (x,t)
 ={1\over 2} \left[  \delta (x-t) + \delta (x+t)\right] \,.
\end{equation}
\vsp
In the case $1/2 <\nu  <1$, the  Green function $\Gc$ will be
determined in the next subsection by using the technique of the Laplace and the Fourier transforms. 
The representations of the  Green function $\Gc$ are of course not new and have been discussed 
in \cite{Mai94}-\cite{Mai11} to mention only a few of  the many papers devoted to this topic.
\vsp
In this paper, we are interested in investigation of some important characteristics of the Green function 
$\Gc$ including location of its maximum point, its propagation speed, and its maximum value.

\subsection{Representations of the Green function}

Following \cite{Mai94}-\cite{Mai11}, some representations of the Green function $\Gc$ 
in form of integrals and series are presented and discussed in this subsection.
\vsp
In \cite{Mai94}, the Laplace and Fourier transforms technique was employed to deduce the 
following representation for the Green function $\Gc$ 
for $x>0$ and $\frac{1}{2}<\nu<1$:
\begin{equation}
\label{recip}
2\nu  \, x\,  \Gc(x,t;\nu  ) 
    = F_\nu(r) = \nu  r\, M_\nu(r),
\end{equation}
where
$$
r={x/t^{\nu  }} >0\,
$$
is the similarity variable and
$$
F_\nu(r) :=
 {\ds {1\over 2\pi i}\,\int_{Ha}}   \!
 \e^{\ds  \sigma -r\sigma ^\nu } \,
   d\sigma \,, \; 
 M_\nu(r) :=
 {\ds {1\over 2\pi i}\,\int_{Ha}}   \!
{\ds \frac{ \e^{\ds  \sigma -r\sigma ^\nu }}{\sigma^{1-\mu}}} \,d\sigma
$$
are the two auxiliary functions nowadays referred to in the  literature of Fractional Calculus as 
the Mainardi functions,
and $Ha$ denotes the Hankel  path properly defined for the representation of the 
reciprocal of the Gamma function.
\vsp
Let us note that the similarity variable $r={x/t^{\nu  }}$ plays a very important role
in our analysis of the location of a maximum point of the Green function $\Gc$.
 In its turn, the form of the similarity variable can be explained by the Lie group analysis
 of the time-fractional diffusion-wave equation (\ref{eq1}). 
\vsp 
 In  \cite{Buc98}and   \cite{Luc98} (see also   \cite{Gor00}),
 symmetry groups of scaling transformations for the time- and space-fractional partial
 differential equations have been constructed. In particular, it has been proved in \cite{Buc98}
 that the only invariant of the symmetry group $T_\lambda$ of scaling transformations of
 the time-fractional diffusion-wave equation (\ref{eq1}) has the form $\eta(x,t,u)={x/t^{\nu}}$
 that explains the form of the scaling variable.
\vsp
Using the well known representation of the Wright function,	which reads (in our notation) for $z\in \CC$
\begin{equation}
\label{wright}
  W_{\lambda ,\mu }(z ) :=  {\ds  {1\over 2\pi i}\,\int_{Ha}}   \!\!
 {\ds \frac{\e^{\, \ds \sigma +z\sigma ^{-\lambda }}}{\sigma^{\mu}} \, d\sigma} = 
   {\ds  \sum_{n=0}^{\infty}{z^n\over n!\, \Gamma(\lambda  n + \mu )}}
   \,,
\end{equation}
where $\lambda >-1$ and $\mu >0$,
we recognize that
the auxiliary functions are  related to the Wright function according
to
\begin{equation}
\label{FM}
{\ds F_\nu(z) =   W_{-\nu , 0}(-z) = \nu  \, z \, M_\nu(z) \,,}    
\quad  {\ds  M_\nu(z) =  W_{-\nu , 1-\nu }(-z)\,.}
\end{equation}
The formula (\ref{FM}) along with (\ref{wright}) provides us with the series representations of 
the Mainardi functions and thus of the Green function $\Gc$ 
(for $x>0$):
\begin{equation}
\label{Gc}
\Gc(x,t;\nu) = {\ds {1\over 2\, \nu\, x}\, F_\nu(r)}
   =	{\ds {1\over 2 \, t^\nu}\, M_\nu(r)}
  =   
{\ds {1\over 2 \, t^\nu}\,\sum_{n=0}^{\infty}\frac{(-x/t^{\nu  })^n}{ n!\, \Gamma(-\nu\, n + 1 -\nu)}}.
\end{equation}
The formulas (\ref{FM})-(\ref{Gc}) can be used to give a new proof of the known fact that 
the support of the Green function $\Gc$ is not compact on the real line for each $t>0$, i.e. 
that a response of the time-fractional diffusion-wave equation with $1/2 < \nu < 1$ to a localized
disturbance spreads infinitely fast. 
Indeed, because the Wright function (\ref{wright}) is an analytical function for $\lambda >-1$ and $\mu >0$
 (see e.g. \cite{Gor99}) that is not identically equal to zero ($W_{\lambda ,\mu }(0 )=1/\Gamma(\mu)>0$), 
 the set of its zeros is discrete and has no finite limit points in the complex plane and thus on the real line. 
 This means that the support of the function $\Gc(x,t;\nu) = W_{-\nu,1-\nu}(-x/t^{\nu  })$ 
 is not compact on the real line for each $t>0$. 
 This fact was proved in \cite{Fuj90} using the representation (\ref{Gc}) of $\Gc$ 
as a function depending on the similarity variable and the asymptotics of this function.
\vsp
Finally we mention another integral representation of the Green function
$\Gc$ that can be found e.g. in \cite{Mai01} or \cite{PovXX}:
\begin{equation}
             \Gc(x,t;\nu)= \frac{1}{\pi}\int_{0}^{\infty}
             E_{2\nu}\left( -\kappa^{2} t^{2\nu} \right)\, \cos (x\kappa)\, d\kappa,
\label{eq11}
\end{equation}
where $E_{\alpha}(z)$ is the Mittag-Leffler function defined by
the series
\begin{equation}
E_{\alpha}(z) = \sum \limits_{n=0}^{\infty} \frac{z^{n}}{\Gamma (\alpha  n+ 1)},
\quad \alpha > 0.
\label{eq12}
\end{equation}
The representation (\ref{eq11}) can be easily obtained by transforming 
the Cauchy problem for the equation (\ref{eq1}) into the Laplace-Fourier domain using the known formula
\begin{equation}
{\cal L}\left\{
\frac{d^{\alpha} u(t)}{dt^{\alpha}};s \right\} =
s^{\alpha}
{\cal L} \left\{
u(t); s \right\} - \sum \limits_{k=0}^{n-1} u^{(k)}(0^{+})s^{\alpha -1-k}, 
\; n-1 < \alpha \le n,
\label{eq3}
\end{equation}
with $ n\in \NN$, for the Laplace transform of the Caputo fractional derivative. 
This formula together with the standard formulas for the Fourier transform of the 
second derivative and of the Dirac $\delta$-function lead to the representation
\begin{equation}
{\widehat{\widetilde{\Gc}}}(\kappa,s,\nu) =
\frac{s^{2\nu -1}}{s^{2\nu} + \kappa^{2}},\ \nu = \alpha/2
\label{eq10}
\end{equation}
of the Laplace-Fourier transform ${\widehat{\widetilde{\Gc}}}$ of the Green function $\Gc$.
Using the well-known Laplace transform formula (see e.g. \cite{Pod99})
$$
{\cal L} \left\{
E_\alpha(-t^\alpha);s \right\}  = \frac{s^{\alpha-1}}{s^\alpha +1}
$$
and applying to the R.H.S of the formula (\ref{eq10}) first the inverse Laplace transform and 
then the inverse Fourier transform we obtain the integral representation (\ref{eq11}) 
if we take into consideration the fact that the Green function of the Cauchy problem is an even function of $x$ 
that follows from the formula (\ref{eq10}).

\subsection{Maximum points of the Green function $\Gc$}

In Fig. 1, several plots of the Green function $\Gc(x;\nu):=\Gc(x,1;\nu)$ 
for different values of the parameter $\nu\ (\nu = \alpha/2)$ are presented 
(see the next section for description of numerical algorithms and programs used to 
calculate the numerical values of the Green function). 
It can be seen that each Green function has an only maximum and that location of the maximum point 
changes with the value of $\nu$. 
\vsp
In Fig. 2, the Green function $\Gc(x,t;\nu)$ is plotted for $\nu = 0.875$ from different perspectives. 
The plots show that both the location of maximum and the maximum value depend on the time $t>0$: 
whereas the maximum value decreases with the time (Fig. 2, right), 
the $x$-coordinate of the maximum location becomes even larger (Fig. 2, left).  
\begin{figure}
\begin{center}
\includegraphics[width=7.25cm]{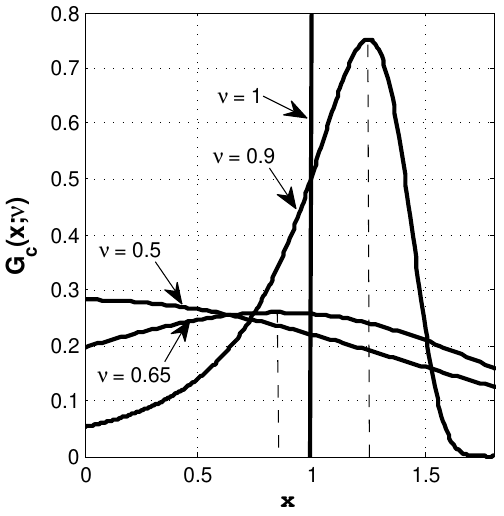}    
\caption{Green function $\Gc(x;\nu):=\Gc(x,1;\nu)$: Plots for several different values of $\nu$}
\label{fig:Green}
\end{center}
\end{figure}
\begin{figure}
\begin{center}
\includegraphics[width=6.3cm]{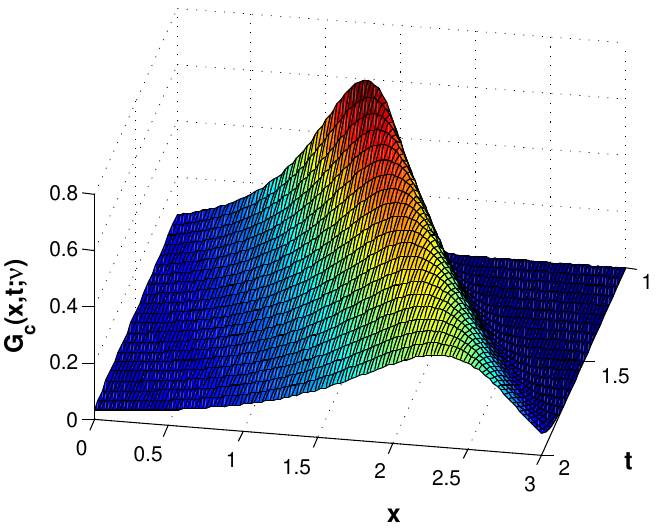} 
\includegraphics[width=6.3cm]{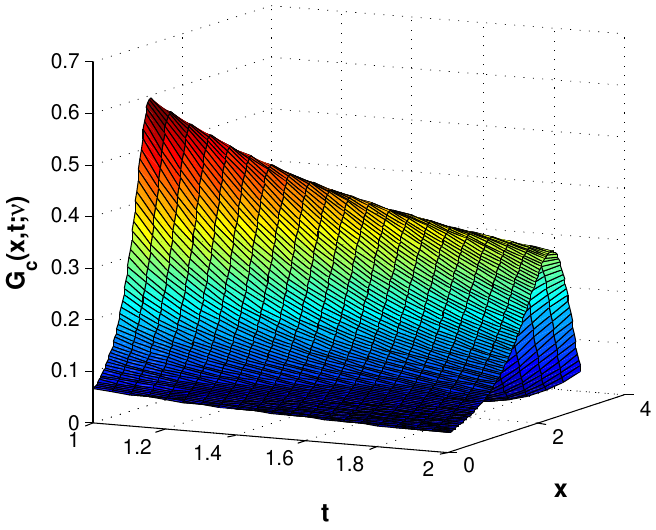} 
\caption{Green function $\Gc(x,t;\nu)$: Plots for $\nu = 0.875$ from different perspectives}
\label{fig:Green3D-x}
\end{center}
\end{figure}
\vsp
The aim of this subsection is to present some analytical formulas that describe both 
the location of the maximum of the Green function $\Gc$, its value, and their interconnection, 
as well as the propagation speed of the maximum location.  
\vsp
In the paper \cite{Fuj90}, an elegant proof of the fact that the Green function $\Gc(x,t;\nu)$ 
of the Cauchy problem takes its maximum at the point 
$x_{*}(t,\nu)=\pm c_\nu t^{\nu}$ for each $t>0$, where $c_\nu >0$ 
is a constant determined by $\nu,\ 1/2<\nu <1$, has been presented. 
His reasoning was as follows: Let us consider the Green function at the point 
$t=1$ and for $x\ge 0$: $\Gc(x;\nu):=\Gc(x,1;\nu)$. 
For $1/2<\nu <1$ the function $\Gc(x;\nu)$ is a stable pdf and the stable pdfs are all unimodal 
(see e.g. \cite{CM61}). 
This means that $\Gc(x;\nu)$  takes its maximum at a certain point $x_{**}=c_\nu$ with a constant 
$c_\nu$ depending on $\nu$. It follows from the formula
(\ref{recip}) that
\begin{equation}
\label{gr1}
\Gc(x;\nu)  = \frac{1}{2} M_\nu(x)
\end{equation}
and
\begin{equation}
\label{gr2}
\Gc(x,t;\nu) =  \frac{t^{-\nu}}{2} M_\nu(xt^{-\nu}) = t^{-\nu}  \Gc(xt^{-\nu};\nu  ).
\end{equation}
Because the function  $\Gc(x;\nu)$  takes its maximum at the point $x_{**}=c_\nu$, the function $\Gc(x,t;\nu)$ has to take its maximum at the point $x_{*}$ that satisfies the relation $x_{*}t^{-\nu} = x_{**}=c_\nu$ due to the formula (\ref{gr2}). Thus the maximum point of the Green function $\Gc(x,t;\nu)$ is moving with the time according to the formula
\begin{equation}
\label{max}
x_{*}(t) = c_\nu t^{\nu},\ \nu = \alpha/2.
\end{equation}
As we see, the main argument in Fujita's proof is dependence of the Green function from the 
similarity variable $xt^{-\nu}$.
\vsp
This argument can be used to give an analytical proof of the relation (\ref{max}) 
following an idea presented in \cite{PovXX}. It should be noted
that derivation of (\ref{max}) given in \cite{PovXX} is based 
on the integral representation (\ref{eq11}) and contains some divergent integrals that 
should be interpreted in one or another generalized sense. 
To avoid this, we present here another proof of  (\ref{max}) 
based on the representation (\ref{Gc}) of the the Green function via the Mainardi function 
and not on the integral representation (\ref{eq11}).
\vsp
Because the Mainardi function $M_\nu$ is an analytical function for $\nu < 1$ as a particular case 
of the Wright function and because of (\ref{Gc}), 
there exist partial derivatives of the Green function 
$\Gc(x,t;\nu)$ of arbitrary orders for $t>0$, $x>0$ 
and we can use the standard analytical method for finding its extremum points. 
We first fix a value $t>0$ and look for the critical points of the Green function $\Gc(x,t;\nu)$ 
that are determined as solutions to the equation
\begin{equation}
\label{der1}
\frac{\partial}{\partial x}\Gc(x,t;\nu)= \frac{t^{-2\nu}}{2} M_\nu^{\prime}(xt^{-\nu})=0
\end{equation}
or to the equation
\begin{equation}
\label{der2}
M_\nu^{\prime}(xt^{-\nu})=0.
\end{equation}
We are interested in a function $x_{*}=x_{*}(t)$ 
that for each $t>0$
defines a solution to the equation (\ref{der2}) . 
The equation (\ref{der2}) can be interpreted as an implicit function that determines the function 
$x_{*}=x_{*}(t)$ we are looking for.
The time-derivative of $x_{*}(t)$ 
can be found with a standard formula for the derivative of an implicit function:
$$
\frac{dx_{*}}{dt} = x_{*}^\prime(t) = 
-\frac{\left.\frac{\partial}{\partial t} M_\nu^{\prime}(xt^{-\nu})\right|_{x=x_{*}}}{\left.\frac{\partial}{\partial x} 
M_\nu^{\prime}(xt^{-\nu})\right|_{x=x_{*}}} 
=\left. -\frac{  -\nu \,t^{-\nu-1}\,x \,M_\nu^{\prime\prime}(xt^{-\nu})}{t^{-\nu}\,
 M_\nu^{\prime\prime}(xt^{-\nu})}\right|_{x=x_{*}} = \nu \frac{x_{*}}{t}.
$$
We thus obtained a simple differential equation for $x_{*}(t)$ with the solution
$x_{*}(t) = C t^{\nu}$, where $C= x_{*}(1)=c_\nu$ 
that is in accordance with the formula (\ref{max}). 
It is known that for $\nu=1/2$ (diffusion equation) $c_\nu = 0$ (the Green function takes its maximum at the point $x=0$ for every $t\ge 0$), whereas for $\nu=1$ (wave equation) $c_\nu = 1$. 
In Section 3, results of the numerical evaluation of $c_\nu,\ 1/2 < \nu < 1$ are presented and discussed. 
\vsp
Let us mention that the same method can be applied for any twice-differentiable 
function that depends on the similarity variable $xt^{-\nu}$ 
or another one in the form of a product of the power functions in $x$ and $t$. 
As it is known, the Green functions for many linear partial differential equations 
of fractional order possess this property and can be investigated by the method presented above. 
In particular, we refer to the recent paper \cite{Luc12}, where the maximum location  of the Green function for the fractional wave equation has been investigated. 
It is worth  mentioning that for the fractional wave equation the constant $c_\nu$ could be determined in analytical form in terms of some elementary functions.
\vsp
As mentioned in \cite{Fuj90}, the maximum point of the Green function $\Gc(x,t;\nu)$ 
propagates for $t>0$ with a finite speed $v(t,\nu)$ that is determined by
\begin{equation}
\label{speed1}
v(t,\nu):=x_{*}^{\prime}(t) = \nu c_\nu t^{\nu-1}.
\end{equation}
This formula shows that for every $\nu,\ 1/2 < \nu < 1$ the propagation speed of the maximum point of 
the Green function $\Gc$ is a decreasing function in $t$ that varies from $+\infty$ at time $t=0+$ 
to zero as $t\to +\infty$. 
For $\nu = 1/2$ (diffusion) the propagation speed is equal to zero because of  $c_{1/2} = 0$ 
whereas for $\nu = 1$ (wave propagation) it remains constant and is equal to $c_{1} = 1$.  
\vsp
\begin{figure}
\begin{center}
\includegraphics[width=8.0cm]{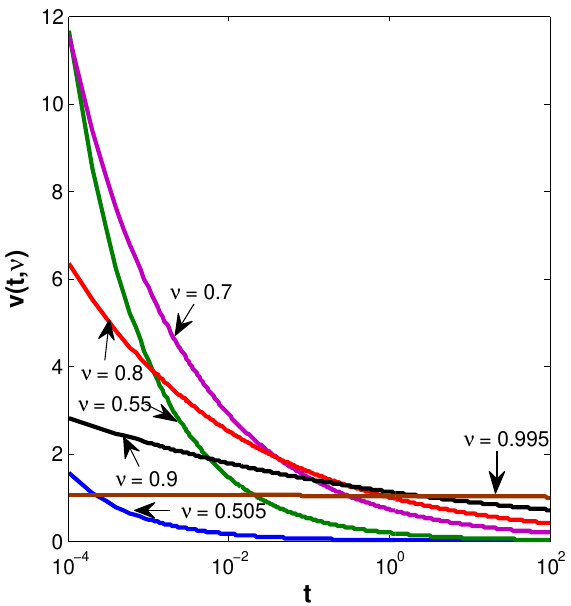}    
\caption{Propagation speed of the maximum point: 
Plot of $v(t,\nu)$ for different values of $\nu$ in the log-lin scale}
\label{fig:speed}
\end{center}
\end{figure}
\vsp
In Fig. 3, some plots of the propagation speed of the maximum point of the Green function 
$\Gc$ are given for different values of $\nu$. 
For large values of $t$,  the smaller the value of $\nu$ is, 
the smaller is the propagation speed for the same time instant. 
Conversely, according to the formula (\ref{speed1}), the smaller the value of $\nu$ is, 
the bigger is the propagation speed for the same time instant when $t\to 0^+$. For example, the propagation speed for $\nu = 0.505$ becomes  greater than the one for $\nu = 0.55$ for $t < 3.04E-24$ (of course, this effect is not visible in the plot of Fig. 3). 
\vsp
Now we determine the maximum value of $\Gc(x,t;\nu)$ in dependence of time.
Let us denote the maximum value by $\Gc^{*}(t;\nu)$ and find it by using the integral representation (\ref{eq11}):
\begin{equation}
\label{MaxVal}
\Gc^{*}(t;\nu):=\Gc(x_{*}(t),t;\nu) = 
\frac{1}{\pi}\int_{0}^{\infty}
             E_{2\nu}\left( -\kappa^{2} t^{2\nu} \right)\, \cos (c_\nu t^{\nu}\kappa)\, d\kappa.
\end{equation}
The variables substitution $\tau = t^{\nu}\kappa$ reduces the integral in (\ref{MaxVal}) to the form
\begin{equation}
\label{MaxVal1}
\Gc^{*}(t;\nu)= \frac{t^{-\nu}}{\pi}\int_{0}^{\infty}
             E_{2\nu}\left( -\tau^{2} \right)\, \cos (c_\nu \tau)\, d\tau,
\end{equation}
i.e. the maximum value $\Gc^{*}(t;\nu)$ of the Green function can be written in the form
\begin{equation}
\label{MaxVal2}
\Gc^{*}(t;\nu)= m_\nu t^{-\nu},
\end{equation}
\begin{equation}
\label{MaxVal3}
m_\nu=\frac{1}{\pi}\int_{0}^{\infty}
             E_{2\nu}\left( -\tau^{2} \right)\, \cos (c_\nu \tau)\, d\tau.
\end{equation}
Moreover, it follows from the formula (\ref{recip}) that
\begin{equation}
\label{MaxVal-n}
\Gc^{*}(t;\nu) = \Gc(x_{*}(t),t;\nu  ) = \frac{1}{2t^\nu}M_\nu(c_\nu)
\end{equation}
with the Mainardi function $M_\nu$, so that we get the relation
$$
m_\nu = \frac{1}{2}M_\nu(c_\nu).
$$
It is well known (see (\ref{diff}) and (\ref{wave})) that $m_\nu = \frac{1}{2\sqrt{\pi}}$ for $\nu = 1/2$ (diffusion equation) and $m_\nu \to +\infty$ as $\nu \to 1$ (wave equation). 
For $1/2 < \nu < 1$, the value of $m_\nu$ can be numerically evaluated 
(see Section 3 for details). 
\vsp
It follows from the relations (\ref{max}) and (\ref{MaxVal2}) or (\ref{MaxVal-n}) that the product 
\begin{equation}
\label{MaxVal4}
\Gc^{*}(t;\nu)\cdot x_{*}(t) = c_\nu\, m_\nu,\ 0<t<\infty
\end{equation}
is a constant that depends only on $\nu$ or on the order $\alpha$ 
of the fractional derivative in the equation (\ref{eq1}), 
i.e., that the maximum locations and the corresponding maximum values specify a 
certain hyperbola for a fixed value of $\alpha$ and for $0<t<\infty$. 
This fact easily follows from the scaling property of the Green function (see (\ref{MaxVal-n})). 
Let us note that the product $\Gc^{*}(t;\nu)\cdot x_{*}(t)$ 
is equal to zero in the case $\nu = 1/2$ (diffusion equation) 
because the maximum point is always located at the point $x_{*}=0$ 
and to infinity in the case $\nu = 1$ (wave equation) 
because the maximum value is always equal to infinity. 
The product values for $1/2 < \nu < 1$ are finite and lying between these extreme values 
that justifies the fact that the time-fractional diffusion-wave equation 
interpolates between the diffusion and the wave equations. 
\vsp
In Fig. 4, we give some plots of the parametric curve ($x_{*}(t),\ \Gc^{*}(t;\nu)$) for
 $0<t<\infty$ that 
is in fact a hyperbola  for different values of $\nu$. 
The vertex of the hyperbola tends to the point $(0,0)$  when $\nu$ tends to $1/2$ (diffusion equation) 
and to infinity when $\nu \to 1$ (wave equation). 
\vsp
\begin{figure}
\begin{center}
\includegraphics[width=7.3cm]{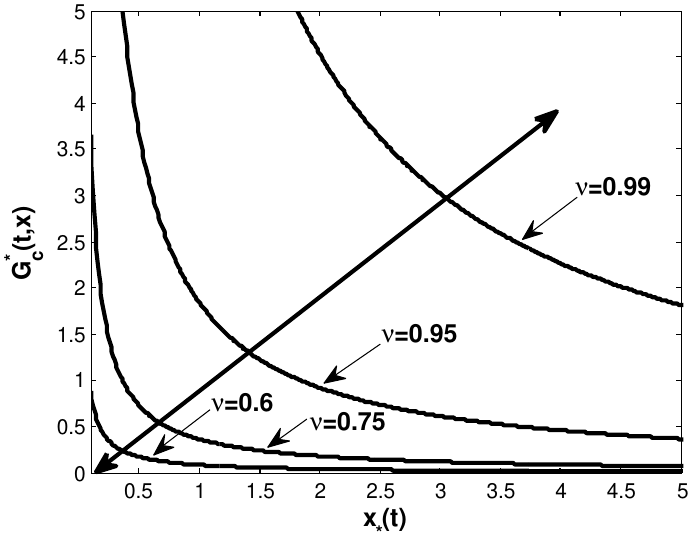}    
\caption{Maximum locations and maximum values of $\Gc(x,t;\nu)$ for a fixed value of $\nu$: 
Plots of the parametric curve ($x_{*}(t),\ \Gc^{*}(t;\nu)$),\ $0<t<\infty$ for different values 
of $\nu$ in the lin-log scale}
\label{fig:hyperbola}
\end{center}
\end{figure}
\vsp
Another interesting and important curve is presented in Fig. 5, 
where the product $c_\nu m_\nu$ of the  maximum location and the maximum value of the Green function 
$\Gc(x;\nu)$ is plotted for $1/2 < \nu <1$.  
As we have seen above, the constants $c_\nu$ (maximum location of $\Gc(x;\nu)$) and 
$m_\nu$ (maximum value of $\Gc(x;\nu)$) are decisive for the behaviour of the Green function 
$\Gc(x,t;\nu)$ for all $t>0$ because the maximum locations and values of this function 
can be determined via these constants for any time point $t>0$ (see the formulas (\ref{max}) and (\ref{MaxVal2})). 
As we can see in Fig. 5, the product $c_\nu\, m_\nu$ is a  monotonically increasing  
function that takes values between $0$ (diffusion equation) and  
$+\infty$ (wave equation). For $0.56 < \nu < 0.99$, 
the product varies between $0.1$ and $10$ , i.e. it changes very slowly on this interval. 
For $\nu \to 1/2$ and $\nu \to 1$ the product $c_\nu\, m_\nu$ goes to $0$ (diffusion equation) 
and to $+\infty$ (wave equation), respectively, very fast. 
\vsp
\begin{figure}\begin{center}
\includegraphics[width=7.0cm]{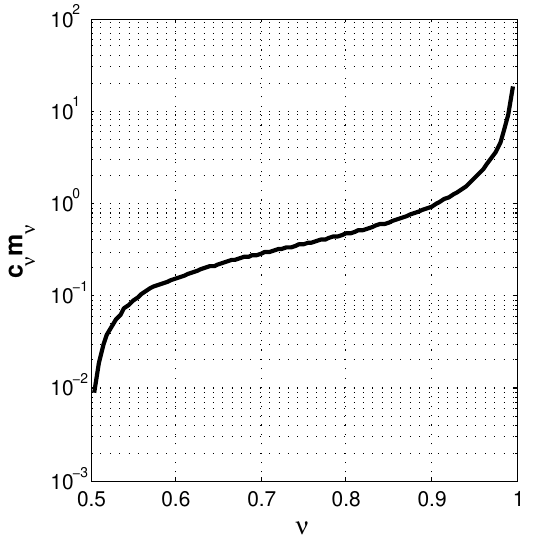}    
\caption{Product of maximum locations and maximum values of $\Gc(x,t;\nu)$ for a fixed time $t=1$: Plot of $c_\nu\, m_\nu,\ 1/2 < \nu < 1$ in the lin-log scale}
\label{fig:Prod}
\end{center}
\end{figure}

\section{Numerical algorithms and results}

In the previous section, some analytical results regarding the location of the maximum of the Green function 
$\Gc$, its maximum value, and the propagation speed of the maximum point as well as 
the plots in Fig. 1 - Fig. 5 were presented. 
Because the analytical formulas derived in the previous section contain the constants $c_\nu$ and $m_\nu$ 
that we could not determine in analytical form, 
we used some numerical algorithms and MATLAB programs for their calculation. 
These algorithms along with some numerical results and plots are presented in this section.
\vsp
To start with, we first discuss algorithms for numerical evaluation of the Green function $\Gc$. 
Because $\Gc$ is a particular case of the Wright function (see formula (\ref{FM})), 
one can of course use the algorithms for the numerical evaluation of the Wright function suggested  
in \cite{Luc08} to evaluate the Green function $\Gc$.
\vsp
Another possibility to calculate the Green function $\Gc$ would be to employ its connection with the stable densities and then to use the existing routines for their numerical calculation (see e.g. \cite{LiCh13} or \cite{N97}). In fact, for $x>0,\ t>0$ the Green function $\Gc$ is connected with the extremal stable density $L_{1/\nu}^{1/\nu-2}$ (see the formulas (4.4) and (4.34) in \cite{Mai01}):
\begin{equation}
\label{stable}
\Gc(x,t;\nu) = \frac{1}{2\nu} t^{-\nu} L_{1/\nu}^{1/\nu-2}\left(\frac{x}{t^{\nu}}\right),
\end{equation}
where $L_\alpha^\theta(x)$ is a stable density (in the Feller parameterization) with the characteristic function given by
$$
\widehat L_\alpha^\theta (\kappa) = \exp\left(-
|\kappa|^\alpha \, \e^{\ds  i (\hbox{sign}\, \kappa)\theta\pi/2}\right),\ 
 0<\alpha  \le 2\,, \ \ 
 |\theta| \le  \,\hbox{min}\, \ \  \{\alpha ,2-\alpha \}.
$$
\vsp 
One more approach to numerical calculation of $\Gc$ that was employed to
 produce our plots for this paper is to
use the integral representation  (\ref{eq11}). 
To calculate the Mittag-Leffler function $E_\alpha$ in (\ref{eq11}), 
we applied the algorithms suggested in \cite{Gor02} and the MATLAB programs that implement these
algorithms and are available from \cite{MC}.
Because the Mittag-Leffler function has for $0< \alpha < 2$ the asymptotics (see e.g \cite{Pod99})
\begin{equation}
\label{MLa}
E_\alpha(-x) = \frac{1}{x\, \Gamma(1-\alpha)}+O(x^{-2}),\ x\to +\infty,
\end{equation}
we can estimate the length of the finite integration interval in the improper integral (\ref{eq11}) 
that allows to reach the desired accuracy $\epsilon$. Indeed, let $A>>t^{-\nu}$  and
$$
\frac{1}{\epsilon} \frac{2t^{-2\nu}}{\pi |\Gamma(1- 2\nu)|} < A.
$$
Then the estimate
$$
|E_{2\nu}(-\kappa^{2} t^{2\nu} )| \le \frac{2}{\kappa^{2} t^{2\nu} \, |\Gamma(1-2\nu)|}
$$
holds true for $\kappa > A$ because of the asymptotic expansion (\ref{MLa}) and we have
$$
\frac{1}{\pi}\left|\int_A^\infty
\!\! E_{2\nu}\left( -\kappa^{2} t^{2\nu} \right)\, \cos (x\kappa)\, d\kappa\right| 
\le \int_A^\infty  \!\! \frac{2}{\pi \kappa^2 t^{2\nu} |\Gamma(1-2\nu)|}d\kappa
=\frac{1}{A} \frac{2t^{-2\nu}}{\pi |\Gamma(1-2\nu)|} <\epsilon.
$$
The integral
$$
\frac{1}{\pi}\int_{0}^{A}
             E_{2\nu}\left( -\kappa^{2} t^{2\nu} \right)\, \cos (x\kappa)\, d\kappa 
$$
with a finite value of $A$ can  then be calculated using any of the known quadrature formulas. 
If $A$ satisfies the conditions mentioned above, we get the estimate 
$$
\left|\frac{1}{\pi}\int_{0}^{A}
             E_{2\nu}\left( -\kappa^{2} t^{2\nu} \right)\, \cos (x\kappa)\, d\kappa - \Gc(x,t;\nu)\right| < \epsilon
$$
with the desired accuracy $\epsilon$ that was used for numerical evaluation of the Green function $\Gc$. 
The results of the numerical evaluation of the Green function $\Gc$ for the time $t=1$ 
and for different values of $\nu$ are presented
in Fig. 1. In Fig. 2, 3D-plots of the Green function are given for $\nu=0.875$, $1\le t\le 2$, 
and $0\le x \le 3$ that illustrate a typical behavior of $\Gc$. 
As can be seen in Fig. 1 and as expected, $\Gc(x,1;\nu)$ has a unique maximum for each $1/2\le \nu \le 1$
and the maximum location changes with $\nu$. Surprisingly,
the maximum location does not always lie between zero (maximum location for the diffusion equation, 
$\nu = 1/2$) and one (maximum location for the wave equation, $\nu =1$). 
Below we consider this phenomenon in more detail.
\vsp
As we have seen in the previous section (formula (\ref{max})), 
the location of the maximum point of the Green function $\Gc(x,t;\nu)$ 
depends on the constant $c_\nu$, i.e. on the location of its maximum for $t=1$. 
It is therefore very important to calculate 
$c_\nu$ numerically and to visualize the dependence of $c_\nu$ on $\nu, 1/2\le \nu \le 1$. 
Because we  already know how to calculate the Green function $\Gc(x,1;\nu)$ 
and because it possesses a unique maximum point $x_{*}=c_\nu$, 
it is an easy task to find the maximum location e.g. with the MATLAB Optimization Toolbox. 
The results of the calculations are presented in Fig.6.
\vsp
Let us note that in \cite{N97} the mode location of the stable densities $f(x;\alpha,\beta)$ 
(in the Nolan parameterization) was numerically calculated and plotted for $0< \alpha \le 2$ and some fixed values of $\beta,\ -1 \le \beta \le 1$. In the case of the Green function $\Gc$, the parameters of the corresponding stable density are connected to each  other (see  (\ref{stable})), 
so that the results presented in Fig.6 are different from ones given in \cite{N97}.
\newpage
\begin{figure}
\begin{center}
\includegraphics[width=5.8cm]{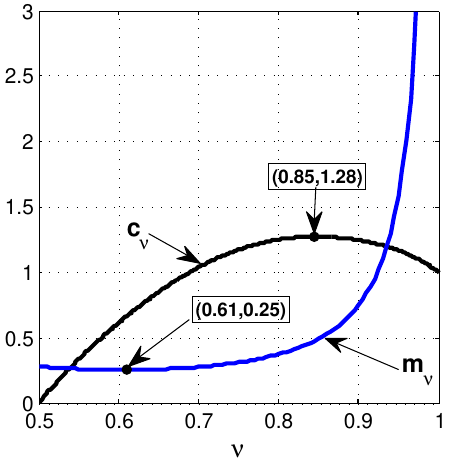}    
\caption{Maximum locations and maximum values of the Green function $\Gc(x;\nu)$: 
Plots of $c_\nu$ and $m_\nu$ for $1/2 \le \nu \le 1$}
\label{fig:MaxLoc}
\end{center}
\end{figure}
\vsp
Fig.6 shows that the curve $c_\nu = c_\nu(\nu)$ has a maximum located at the point $\nu \approx 0.85$. 
The value of the maximum is approximately equal to $1.28$. 
It is interesting to note that for $0.69 \le \nu \le 1$ the value of $c_\nu$ is greater than or equal to one. 
In Fig.6  we also present results of numerical evaluation of the maximum value $m_\nu$ of the Green function  $\Gc(x,t;\nu)$ 
at the time $t=1$ as function of $\nu, 1/2\le\nu <1$. 
It follows from (\ref{MaxVal2}) that the constant $m_\nu$ determines the maximum value 
of $\Gc(x,t;\nu)$ at any time $t>0$.  
For numerical calculation of $m_\nu$, the formula (\ref{MaxVal3}) was used. 
As expected, $m_\nu$ tends to infinity as $\nu$ tends to 1 that corresponds to the case of the wave equation.
Another interesting feature of the curve $m_\nu = m_\mu(\nu)$ that can be seen in Fig.6 
is that $m_\nu$ is first monotonically decreasing and then starts to increase. 
The minimum location of $m_\nu = m_\mu(\nu)$ is at  $\nu \approx 0.61$ 
and the minimum value is nearly equal to $0.25$. 
Whereas $m_\nu$ changes very slowly on the interval $0\le \nu < 0.95$, 
it starts to rapidly grow in a small neighborhood of the point $\nu = 1$. 
It should be noted that despite of the fact that the curves $m_\nu = m_\nu(\nu)$ 
and $c_\nu = c_\nu(\nu)$ are not monotone and possess a minimum and a maximum, respectively, 
the product $c_\nu\, m_\nu$ is a monotone increasing function for all $\nu, 1/2 \le \nu \le 1$ (see Fig. 5). 
\newpage
\subsection*{Acknowlwdgements}
The first named author is grateful to National Institute of Nuclear Physics
(INFN) of Italy for financial support of his visit to the University of Bologna in December 2011.
The authors appreciate constructive remarks and suggestions of the referees that helped to improve the manuscript. 


\end{document}